# B1-B2 phase transition in MgO at ultra-high static pressure

Natalia Dubrovinskaia[1,*], Sylvain Petitgirard[2], Stella Chariton[2], Rémi Tucoulou[3], Jan Garrevoet[4], Konstantin Glazyrin[4], Hanns-Peter Liermann[4], Vitali B. Prakapenka[5], and Leonid Dubrovinsky[2,*]

[1]*Material Physics and Technology at Extreme Conditions, Laboratory of Crystallography, University of Bayreuth, D-95440 Bayreuth, Germany*

[2]*Bayerisches Geoinstitut, University of Bayreuth, D-95440 Bayreuth, Germany*

[3]*European Synchrotron Radiation Facility, BP 220 F-38043 Grenoble Cedex, France*

[4]*Photon Science, Deutsches Elektronen-Synchrotron, Notkestrasse 85, D-22607 Hamburg, Germany*

[5]*Center for Advanced Radiation Sources, University of Chicago, Chicago, Illinois 60437, USA*

*natalia.dubrovinskaia@uni-bayreuth.de and leonid.dubrovinsky@uni-bayreuth.de

**Studies of the behaviour of solids at ultra-high pressures, those beyond 200 GPa, contribute to our fundamental understanding of materials' properties and allow an insight into the processes happening at such extreme conditions relevant for terrestrial and extra-terrestrial bodies. The behaviour of magnesium oxide, MgO, is of a particular importance, as it is believed to be a major phase in the Earth's lower mantle and the interior of super-Earth planets. Here we report the results of studies of MgO at ultra-high static pressures up to ca. 660 GPa using the double-stage diamond anvil cell technique with synchrotron X-ray diffraction. We observed the B1-B2 phase transition in the pressure interval from 429(10) GPa to 562(10) GPa setting an unambiguous reference mark for the B1-B2 transition in MgO at room temperature. Our observations allow constraining theoretical predictions and results of available so far dynamic compression experiments.**

Ferropericlase (Mg,Fe)O is the second most abundant component of the Earth's lower mantle and believed to be a very important phase in the interior of super-Earth planets [1]. The end-member of this solid solution, magnesium oxide MgO, is one of the most studied compounds at high pressure (HP). At normal pressure, this simple ionic oxide has the cubic B1 (NaCl-type) crystal structure, which is stable at room temperature (RT) to at least 227 GPa as revealed by static-compression experiments in the diamond anvil cell (DAC) [2-4]. Due to the high symmetry and RT structural stability, MgO is widely used as a pressure calibration standard in large-volume press and diamond anvil cell experiments.

It has been theoretically predicted that at high-pressure and high-temperature (HPHT) conditions MgO should undergo a phase transition into the HP phase with the B2 (CsCl-type) structure [5-7]. According to the theory, at room temperature the transition should happen at about 500 GPa [5-7]. The different theoretical calculations agree well regarding the slope of the B1-/ B2-MgO solid-solid phase transition boundary [5-8] in the range of temperatures up to 10000 K and pressures between 300 GPa and 500 GPa. However, so far there are no sufficient experimental evidences, which could help to constrain the room-temperature transition point and the slope of the two-phase curve.

Currently, there are substantial discrepancies regarding the position of the B1-/ B2-MgO phase transition boundary, as suggested by dynamic-compression experimental results [9,10] and the one predicted by *ab initio* calculations [5,7]. A solid-solid phase transformation can be unambiguously confirmed only on the basis of experimentally obtained structural information. Unfortunately, shock-compression experiments do not provide any structural information and the report on the B1-B2 phase transition at ca. 440(±80) GPa and 9000(±700) K was based on a temperature anomaly solely [9]. According to McWilliams et al. [9], the value of the Clapeyron slope ($d$P/$d$T) is equal to -3.9(±3.0) x$10^{-4}$ TPa/K, and has a rather large uncertainty. Such a value extrapolated linearly leads to an estimation of the phase transition pressure between 1.2 TPa and 6.4 TPa at room temperature. This estimation disagrees with the theoretical predictions of the slope of the phase boundary and the RT transition point of ca. 0.5 TPa [5-7]. In the ramp-compression study by Coppari et al. [10], the experimental temperature could be constrained neither using classic pyrometry because of the opaqueness of the diamond window, nor using the density measurement and the extrapolation of the thermal expansivity of MgO because of too large uncertainties associated to the measurements [10]. Nevertheless, using dynamic X-ray diffraction (XRD) measurements, a solid-solid phase transition, consistent with a transformation to the B2 structure, was reported near 600 GPa based on the disappearance/appearance of a single peak in the XRD [10]. Finally, laser-shock compression experiments [8] reported that the B1-MgO phase is stable up to at least 350 GPa.

So far, static-compression X-ray diffraction experiments have been conducted at pressures up to 227 GPa [2-4], which are far below the expected B1-/B2-MgO phase transition. Thus, reliable X-ray diffraction data for MgO beyond 250 GPa could contribute to establishing the missing anchor pressure point of the expected B1-/B2-MgO phase transition at room temperature and to reducing uncertainties in the equation of state of MgO. This makes static-compression experiments at ultra-high pressures highly desirable.

Here we report the results of static-compression experiments on MgO in the pressure range up to ~660 GPa using the double-stage diamond anvil cell (dsDAC) technique [11-13]. The HP behaviour of MgO was studied using synchrotron X-ray diffraction. We observed a phase transition from the B1 to B2 crystal structure with a transition point at room temperature between 429(10) and 562(10) GPa. The new data allow a direct comparison between the results of static- and dynamic-compression experiments and theoretical predictions.

*In situ* high-pressure synchrotron X-ray diffraction experiments were conducted at ID16B ESRF (France), P06 PETRA III (Germany), and GSECARS (Sector 13) at APS (United States) (*Methods*). A sample for an experiment is prepared as follows: (1) A rhenium foil (gasket) with an initial thickness of ~200 μm is indented to ~25μm using a conventional DAC with diamond anvils culets of 250 μm. (2) In the middle part of the initial indentation, an additional indentation with a final thickness of ~3 μm is made using 100-μm culet diamonds. (3) Using a pulsed laser, a hole with a diameter of ~3 to 4 μm is drilled at the center of the secondary indentation. (4) A sample made of a mixture of MgO powder and W or Au powder, used as pressure gauge, is loaded using the micromanipulator (MicroSupport Co., Ltd., Japan) into the center of the secondary pressure chamber. Semi-balls made of nanocrystalline diamond (NCD) [13,14], which serve as secondary anvils, are attached to the gasket using traces of paraffin wax, and the whole assembly is mounted on the primary anvils (with either flat 250-μm or bevelled 120-μm culets in our experiments). Empty space in the primary chamber is filled either by Ne under pressure of 1.2 kbar (in one experiment with primary anvils with 250 μm anvils), or by paraffin wax. The dsDAC preparation technique is described in detail elsewhere [11-13]. A schematic of the dsDAC is provided in Fig S1.

In order to verify the consistency of dsDACs results to those obtained from conventional DACs, we conducted an experiment in a dsDAC at the GSECARS beam-line at the APS (with a focused X-ray beam size of ~2 × 3 μm$^2$ FWHM in diameter, see *Methods*) starting from the lowest pressure achievable with a dsDAC. The primary chamber was filled with neon as a pressure-transmitting medium (PTM). Figure S2 shows the diffraction pattern of MgO compressed in a gasketed dsDAC. The XRD appears to be very informative, as it presents the state of materials in the whole dsDAC assembly. Indeed, the full profile analysis using the GSAS program [15,16], reveals the following information: B1-MgO (diffraction lines are marked with black ticks under the diffraction pattern in Fig. S2) in the secondary pressure chamber is under a pressure of 213(10) GPa according to the B1-MgO EOS [2]; Re of the gasket (purple ticks) is under a pressure of 50(3) GPa, according to the Re EOS [11]; Ne PTM is at 41(1) GPa (green ticks) according to the Ne EOS [17]. The pressure revealed by Re diffraction is slightly higher than the one from Ne, although almost within the uncertainty of measurements. However, this value from Re may also reflect an additional load contribution from the gasket compressed by the secondary anvils. Especially remarkable is that the diffraction pattern in Fig. S2 also shows the diffraction lines from different parts of the nanocrystalline diamond NCD anvils (designated as "D-I" and "D-II" in Figs. S2 and S3): "D-I" corresponds to the NCD at 64(5) GPa (red ticks in Fig. S2), which is moderately higher than the pressure recorded from the PTM, and comes from the main body of the NCD anvil. "D-II" (orange ticks in Fig. S2) corresponds to the NCD at 204(10) GPa, which is similar to the pressure in the secondary chamber (213(10) GPa on MgO). The latter diffraction peaks come from the apexes of the NCD anvils. Figure S3 confirms that these observations are in agreement with the design of dsDAC experiments because we could not find evidence of either NCD, or MgO diffraction two microns away from the center of the sample, and only the diffraction lines of Re and Ne are visible in the diffraction pattern (bottom pattern in Fig. S3). Although this experiment did not allow us to reach ultra-high pressures because of the cell failure, it was of a methodological significance for better understanding and further development of the dsDAC technique. It also confirmed that data regarding pressure extracted from static compression experiments with conventional DACs [2] are in agreement with values measured in dsDAC.

Ultra-high pressure experiments in ds-DACs require a high-energy monochromatic X-ray beam with an extremely small size, desirably below a micrometre FWHM in diameter. At the ESRF, ID16B offers the possibility to focus the beam down to (nominally) 70 nm FWHM with a monochromatic energy of 29.6 keV (see *Methods*). However, due to relatively large tails, the effective size of the X-ray beam appears to be significantly larger. For the dsDAC experiments it means that we still observe strong diffraction from the Re gasket, even if we focus the beam in the center of the hole of about ~5 μm in diameter. Nevertheless, with the nano-size beam the signal intensity from the sample is quite comparable to that of the surrounding gasket improving the quality of the diffraction if compared to a micron beam size.

A dsDAC with MgO and W in the secondary pressure chamber (with a starting diameter of about 4 μm and a thickness of about 2.5 μm) was prepared for ultra-HP experiments at ID16B. The secondary anvils were made from NCD balls of ~10 μm in diameter, which were milled using the Focused Ion Beam (FIB) technique in order to obtain two semi-balls [13]. Primary anvils of 80-μm culets bevelled at 300 μm were used, and the primary chamber was of about 40 μm in diameter. The sample was pressurized to a maximum pressure of 562(10) GPa, which was determined from the equation of state of tungsten [12].

Figure 1 shows a sequence of the selected diffraction patterns obtained in the pressure range of 341(10) to 562(10) GPa. The pressure within the secondary pressure chamber was determined using the EOS of W [12]. It shows a consistent increase upon compression. Examples of the diffraction patterns for all pressure points obtained in this experiment are shown in Fig. S4 (details for the pressure determination are also provided in Fig. S5). All diffraction patterns contain reflections of the sample (MgO), pressure marker (W) and the gasket material (Re) of the secondary gasket. The highest-pressure point, at which the diffraction of B1-MgO was still observed, was 429(10) GPa. At the next pressure point of 562(10) GPa, the diffraction from MgO is consistent with the B2-MgO crystal structure, although the intensity of the (110) and (111) reflections is invert (i.e. the latter is higher than the former), which can be related to a preferred orientation of the MgO sample in the pressure chamber, as we can identify a few arches and not complete diffraction rings.

A subsequent experiment was conducted at PETRA III (see *Methods*) and aimed at confirming the reproducibility of the observation of the B2-MgO. We loaded MgO powder mixed with Au particles as the pressure marker. The dsDAC was prepared as described above and pre-compressed to about 160 GPa at the Bayerisches Geounstitut (BGI). Pressure was estimated from Raman spectra of the primary anvil. Then, the cell was transferred to P06 beam-line and the sample was investigated using X-ray beam with the size of 0.7x0.7 μm$^2$ FWHM and the wavelength of 0.5900 Å. At the beamline pressure was determined using the Au EOS according to [12]. Figure 2 shows two diffraction patterns obtained at 637(10) GPa (from the cell as-prepared) and 661(10) GPa (after the pressure increase) (Figs. S6 and S7), i.e. above 562(19) GPa, at which we first observed the B2-MgO in the previous experiment described above. For both pressure points measured at P06, the (100) and (110) reflections characteristic for B2-MgO were present. Figure S6 highlights the importance of the small X-ray beam size at a focal spot and the precise alignment of the cell in experiments with dsDACs. The diffraction patterns taken more than one micron away from the center of the sample do not show the material under investigation at all.

Our experiments provided us three P-V points for the B1-MgO phase at 341(4) GPa, 380(7) GPa and 429(10) GPa (Figure 3; red inverted empty triangles), thus considerably extending the pressure range of the previous static-compression experiments conducted below 250 GPa [2-4] using conventional DACs. These three points are in perfect agreement with the B1-MgO EOS of [2] (the black continuous line in Figure 3), which can be thus extended to at least 429 GPa.

From our two experiments at ultra-high pressures above the B1-B2 MgO transformation, we can now add three experimental points in the P-V diagram of MgO for the B2-MgO phase (Figure 3; red triangles) ranging from 562(10) GPa to 661(10) GPa. Obviously, this is not sufficient for an accurate determination of the EOS of B2-MgO. Still, we can constrain our B2-MgO data through a 2$^{nd}$ order Birch-Murnaghan EOS, with a reference volume $V_{562GPa}$=5.36 cm$^3$/mole, and a bulk modulus $K_{562GPa}$=1801(6) GPa (Fig. 3).

Figure 3 also compares our experimental P-V results for B1-MgO and B2-MgO with other currently available experimental and theoretical data. As mentioned before, our experimental data for B1-MgO are in a perfect agreement with the experimental EOS of Duffy et al. [2] and with the theoretical EOS of Cebulla and Redmer [7]. The semi-empirical B1-MgO EOS of Dorogokupets and Dewaele [18] is, somehow, quite different at pressures higher than 175 GPa. Our experimental room temperature B2-MgO EOS is so far the only one available. In comparison with our static compression experiments, the theoretical P-V data of Oganov et al. [5] predict lower compressibility, while those of Cebulla and Redmer [7] – a higher compressibility for the B2-MgO phase.

Figure 4 summarizes all available data on the P-T diagram of the phase relations of MgO. Theoretical data of Oganov et al. [5] and Cebulla and Redmer [7] give quite similar results for the location of the two-phase boundary. Their theoretical pressure points for the B1-B2 phase transition at room temperature fall within the interval between 429 and 562 GPa, in agreement with our results from static compression experiments. The slope of the phase boundary, which can be drawn on the basis of the results of shock-compression experiments by McWilliams et al. [9] (violet, almost horizontal line in Fig. 4), is not in agreement with theoretical predictions. A linear extrapolation of this line suggests a much higher pressure value for the RT B1-/B2-MgO phase transition than any other data from static-compression experiments or theoretical calculations [5-7].

Coppari et al. [10] reported an estimated temperature of 3900±2000 K in their ramp-compression to 600 GPa, which, according to the authors [10], was deduced from the Clapeyron slope ($dP/dT$=-3.9(±3.0) x$10^{-4}$ TPa/K) reported in McWilliams et al. [9]. However, the value of $dT/dP$= -26±30 K/GPa, which Coppari and co-authors used for the temperature estimation (see Suppl. Information in [10], page 10) is by one order of magnitude larger than the value of $dT/dP$= -2.6 K/GPa, which is deduced from the Clapeyron slope ($dP/dT$=-3.9(±3.0) x$10^{-4}$ TPa/K) reported in McWilliams et al. [9]. If the value of $dT/dP$= -2.6 K/GPa is applied, the estimated temperature at 600 GPa in Coppari et al. [10] turns out to be 8600 K, which is similar to the shock-compression temperature reported by McWilliams et al. [9]. The latter value, however, contradicts the statement of Coppari et al. [10] that "temperatures reached in ramp loading are expected to be lower than those obtained along the shock Hugoniot" (see Suppl. Information in [10], page 9). In Fig. 4 the transition point for the ramp-compression [10] is put according to the corrected temperature.

Although dynamic-compression experiments have a very large uncertainty in pressure and temperature, their results are very important, as they contribute to understanding the studied phenomena as a whole. In particular, the dynamic X-ray diffraction measurements [10] appear to be consistent with the results of static-compression experiments with regard to constraining the pressure interval of the B1-/B2-MgO phase transition. At pressures above 600 GPa Coppari et al. [10] observed a diffraction peak, whose shift as a function of pressure was consistent with the equation of state of B2-MgO [10]. At pressures below 420 GPa they observed the diffraction from B1-MgO. Thus, although the temperature was unconstrained, the pressure interval of the transition could be defined as between 420 GPa to 600 GPa, and thus in agreement with our static compression experiments at RT.

To conclude, the present work contributes to the fundamental knowledge of the behaviour of MgO at ultra-high static pressures corresponding to those in the mantles of super-Earth planets. The experimental data on the behaviour of MgO obtained under static compression set an unambiguous reference mark for the B1-B2 phase transition at room temperature. These data could be directly compared to the results of dynamic shock- and ramp-compression experiments and theoretical calculations. The experimental information collected in the course of this work due to the first use of a nano-size X-ray beam, including the pressure/stress distribution in the pressure chamber, serves for the further methodological development of the ds-DAC technique for studies of the behaviour of matter at ultra-high static pressures.

**Methods**

**Sample preparation in ds-DACs.** The ds-DACs [11-13] were prepared in Bayreuth, Germany, using either the BX90 DAC [19] (for experiments at the P06 beamline, Petra III, Hamburg), or a small version of the BX90 (weight of about 180g) to fit the setup on the synchrotron micro-focus beamline ID16B at the ESRF, Grenoble. The methodology of the gasketed ds-DACs preparation is described in detail elsewhere [13]. It enables us to place a small metallic rhenium Re gasket with a tiny hole between two anvils of the second stage - two nano-diamond semi-balls [14] mounted on primary diamond anvils of the BX90 [13]. A sample of MgO (99.99% purity, Sigma Aldrich Inc.) powder is placed into the gasket hole to be pressurized between the second-stage anvils. The size of the sample chamber is of about 3 to 4 μm in diameter and about 2.5 μm in initial thickness. In different experiments either tungsten W or gold Au (both 99.999% purity, Goodfellow Inc) were used as pressure calibrants. Sources of possible contaminations of the studied samples were carefully analyzed; impurities, which could affect the diffraction patterns, were never detected. In all experiments, except those conducted at the APS, paraffin wax was used as a pressure transmitting medium in the primary chamber of the ds-DACs. At the APS, experiments were conducted at pressures below 213 GPa and Ne in the primary pressure chamber was used as a pressure medium.

**Synchrotron X-ray diffraction experiments.** At the ID16B at ESRF, the data were collected with a Frelon CCD taper camera placed at the distance of *circa* 110 mm using the X-ray beam of 0.4824 Å in wavelength and the beam size down to $75 \times 60$ nm$^2$. At GSECARS (APS), experiments were performed utilizing PILATUS3 X CdTe 1M area detector and a tightly focused X-ray beam (~$2 \times 3$ μm$^2$) of 0.2952 Å. We used high resolution highly sensitive optical system available at GSECARS for precise alignment of a sample with the X-ray beam visualized with high-energy X-ray induced florescence on the sample in the dsDAC [20]. At the micro focusing beamline P06 at PETRA III, the data were collected with the Mar165 detector, using the X-ray beam with the wavelength of ~0.5900 Å and the beam size down to 0.7 × 0.7 μm$^2$ using KB mirror systems. Motor sample stack of P02.2 beamline of PETRA III was used in the latter experiment.

**XRD data analysis.** The collected images were integrated using the FIT2D and Dioptas [21] programs to obtain a conventional diffraction pattern. Data analysis was conducted using the GSAS (general structure analysis system) package [15,16].

**Acknowledgements**

N.D. and L.D. thank the Deutsche Forschungsgemeinschaft (DFG projects DU 954-11/1 and DU 393-10/1) and the Federal Ministry of Education and Research, Germany (BMBF, grant no. 5K16WC1) for financial support. Portions of this work were performed at GeoSoilEnviroCARS (The University of Chicago, Sector 13), Advanced Photon Source (APS), Argonne National Laboratory. GeoSoilEnviroCARS is supported by the National Science Foundation - Earth Sciences (EAR - 1634415) and Department of Energy-GeoSciences (DE-FG02-94ER14466). This research used resources of the Advanced Photon Source, a U.S. Department of Energy (DOE) Office of Science User Facility operated for the DOE Office of Science by Argonne National Laboratory under Contract No. DE-AC02-06CH11357.
.

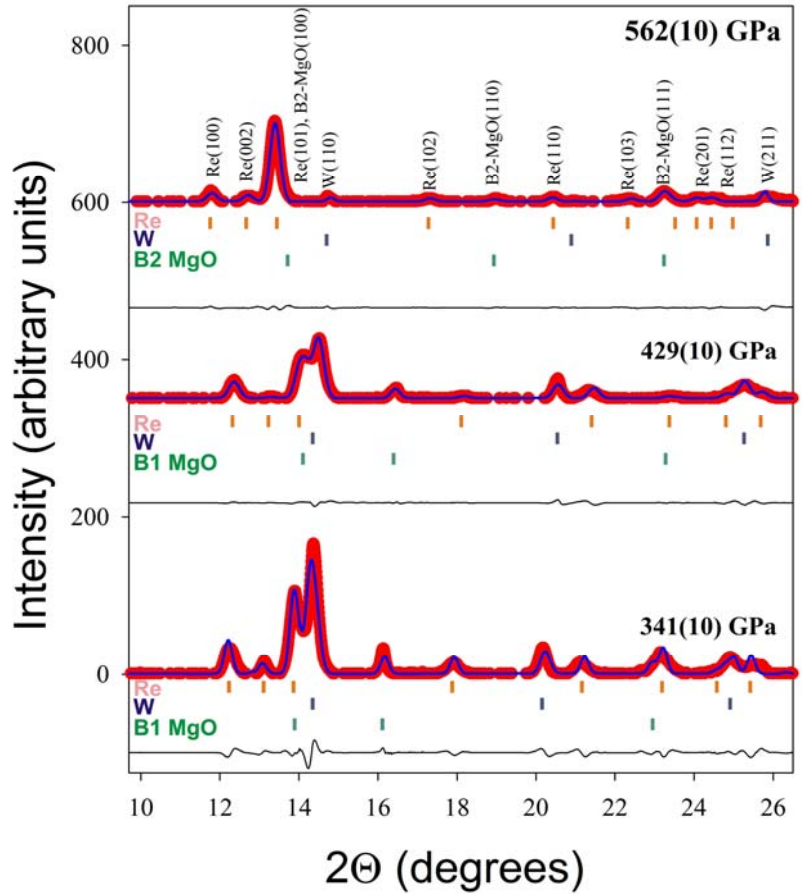

**Fig. 1. Diffraction patterns of MgO and W compressed in a dsDAC in experiments at ID16B, ESRF.** Tungsten was used as a pressure marker. Re peaks are from the gasket. Lower curve: mixture of B1 MgO ($a$=3.4303(7) Å) and W ($a$=2.7484(4) Å)) at 341(10) GPa; Re secondary gasket ($a$=2.6186(4) Å, $c$= 4.2359(12) Å). Middle curve: B1 MgO ($a$= 3.3742(5) Å) and W ($a$= 2.7032(3) Å) at 429(10) GPa; Re secondary gasket ($a$=2.5891(5) Å, $c$= 4.165(3) Å). Upper curve: B2 MgO ($a$=2.0724(4) Å) and W ($a$= 2.6478(4) Å) at 562(10) GPa; Re secondary gasket ($a$=2.7405(4) Å, $c$= 4.4053(14) Å). Experimental data are shown by red dots; continuous blue curves are simulations using the full-profile (GSAS) software [15,16]. "2θ" is the diffraction angle; the labels above the peaks indicate the indices of the diffraction reflections of the corresponding material; the pressures given above the curves designate at which pressure in the dsDAC the diffraction patterns were collected. X-ray wavelength is 0.4824 Å.

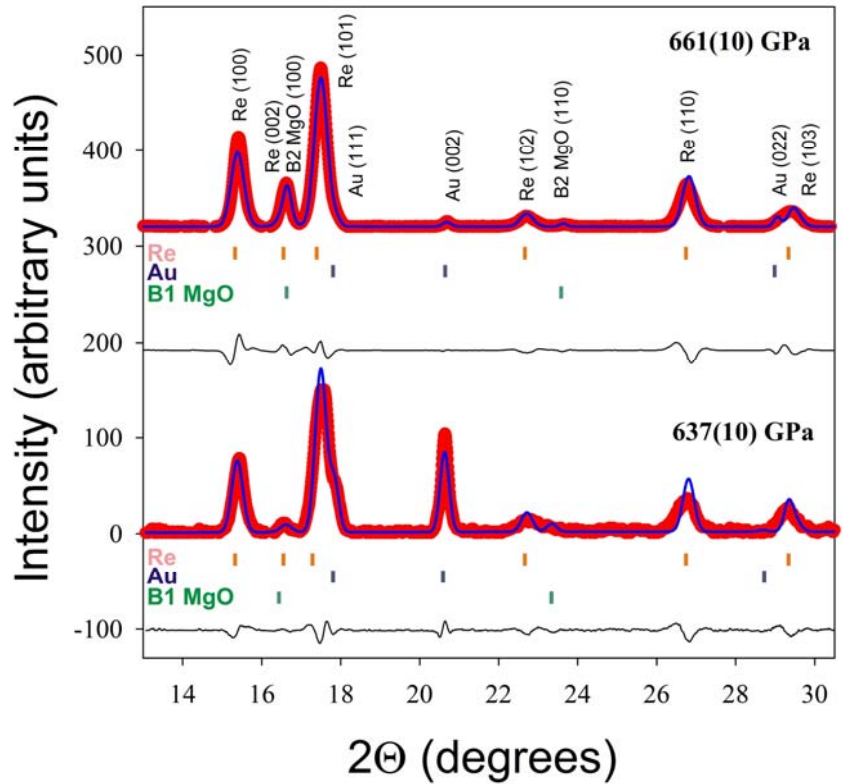

**Fig. 2. Diffraction patterns of MgO and Au compressed in a dsDAC in experiments at P06, PETRA III.** Au was used as a pressure marker. Re peaks are from the gasket. Lower curve: mixture of B2 MgO (*a*= 2.062(1) Å) and Au (*a*= 3.2951(4) Å) at 637(10) GPa; Re gasket (*a*= 2.5452(3) Å, *c*= 4.088(2) Å). Upper curve: B2 MgO (*a*= 2.0379(3) Å) and Au (*a*= 3.2864(9) Å) at 661(10) GPa; Re gasket (*a*= 2.5441(2) Å, *c*= 4.087(1) Å). Experimental data are shown by red dots; continuous blue curves are simulations using the full-profile (GSAS) software [15,16]. "2θ" is the diffraction angle; the labels on the peaks indicate the indices of the diffraction reflection of the corresponding materials; the pressures given above the curves designate at which pressure in the dsDAC the diffraction patterns were collected. X-ray wavelength is 0.5900 Å.

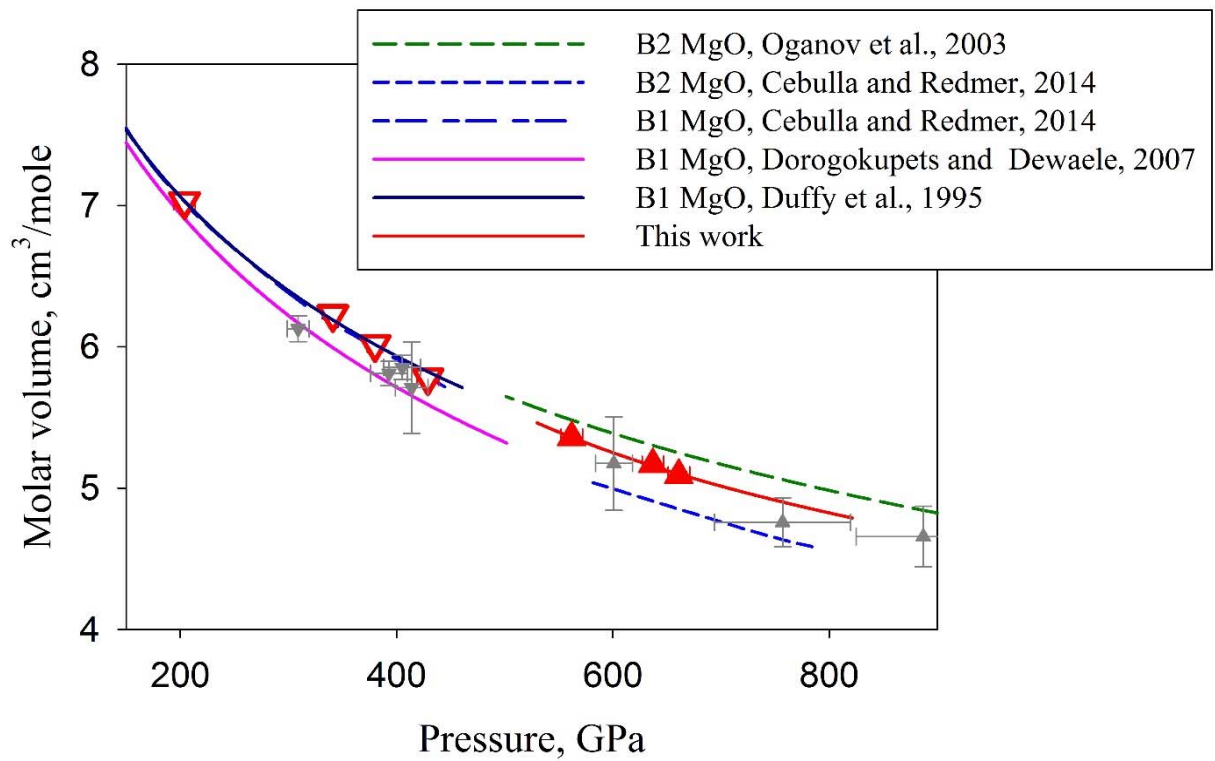

**Fig. 3. The pressure dependence of the molar volume of MgO as determined in experiments and suggested by theory.** Static compression experiments at ambient temperature in this work: solid red triangles – B2-MgO phase, open red inverse triangles – B1-MgO phase; dynamic compression experiments at high temperatures by Coppari et al. [10]: grey triangles – B2 phase, grey inverse triangles – B1 phase. Continues black line is for the B1-MgO experimental EOS by Duffy et al. [2]; continues magenta line is for the B1-MgO semi-empirical EOS by Dorogokupets and Dewaele [18]; dashed blue lines are for the B1- and B2-MgO *ab initio* EOSes by Cebulla and Redmer [7]; dashed green line is for the B2 MgO *ab initio* EOS by Oganov et al. [5]. The continues red line is the fit of the experimental data for B2-MgO from this work with the 2$^{nd}$ order Birch-Murnaghan EOS ($V_{562GPa}$=5.36 cm$^3$/mole, $K_{562GPa}$=1801(6) GPa).

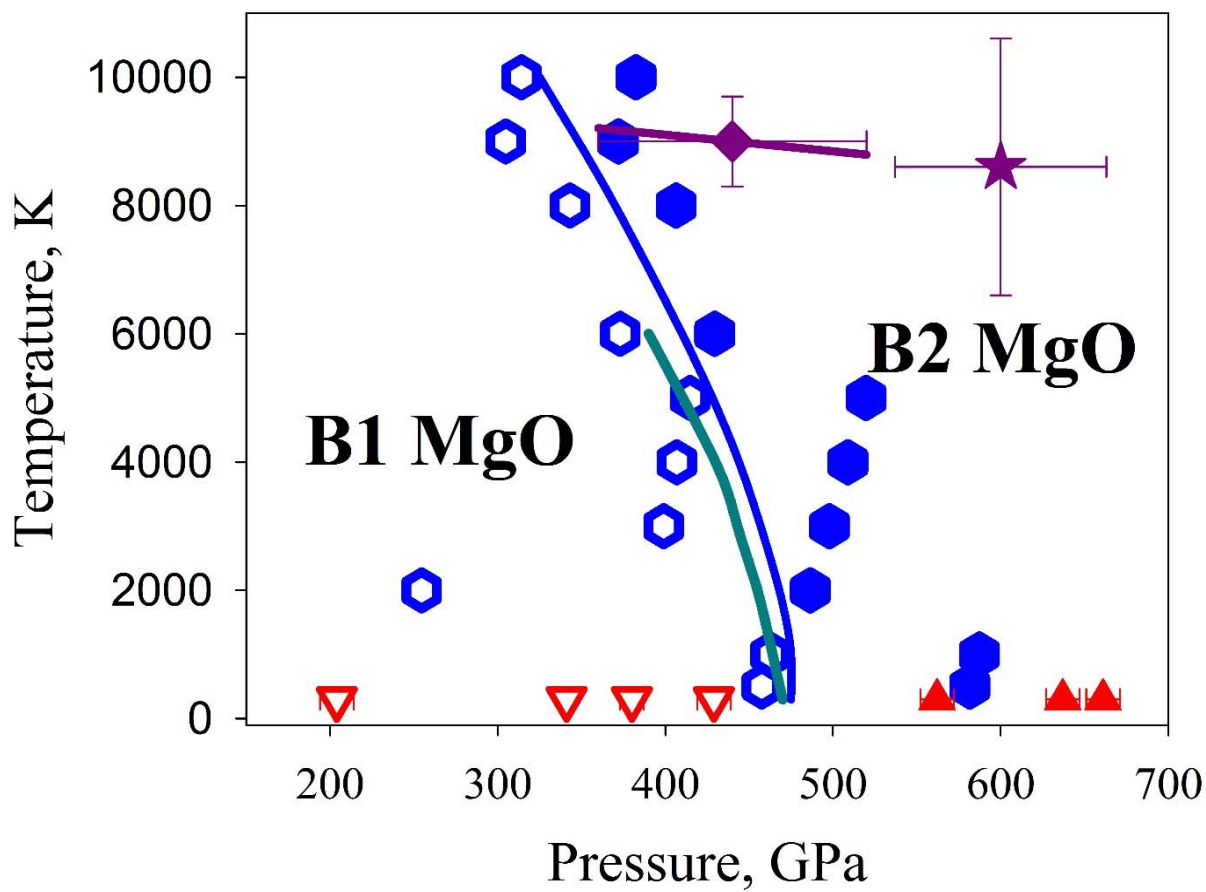

**Fig. 4. The schematic PT diagram of the phase relations for MgO.** Experimental static compression data obtained in this work are designated by solid red triangles (B2-MgO phase), and open red inverse triangles (B1-MgO phase). Dynamic compression data are according to McWilliams et al. (2012) [9] (violet diamond) and Coppari et al. (2013) [10] (violet star) with corrected temperature (see text). Theoretical calculations are according to Oganov et al. (2003) [5] (green continuous line) and Cebulla and Redmer (2014) [7] (blue line and blue hexagons (solid hexagons – B2-MgO, open hexagons – B1-MgO).

# B1-B2 phase transition in MgO at ultra-high static pressure

Natalia Dubrovinskaia[1,*], Sylvain Petitgirard[2], Stella Chariton[2], Rémi Tucoulou[3], Jan Garrevoet[4], Konstantin Glazyrin[4], Hanns-Peter Liermann[4], Vitali B. Prakapenka[5], and Leonid Dubrovinsky[2,*]

[1]*Material Physics and Technology at Extreme Conditions, Laboratory of Crystallography, University of Bayreuth, D-95440 Bayreuth, Germany*

[2]*Bayerisches Geoinstitut, University of Bayreuth, D-95440 Bayreuth, Germany*

[3]*European Synchrotron Radiation Facility, BP 220 F-38043 Grenoble Cedex, France*

[4]*Photon Science, Deutsches Elektronen-Synchrotron, Notkestrasse 85, D-22607 Hamburg, Germany*

[5]*Center for Advanced Radiation Sources, University of Chicago, Chicago, Illinois 60437, USA*

*natalia.dubrovinskaia@uni-bayreuth.de and leonid.dubrovinsky@uni-bayreuth.de

**Supplementary figures**

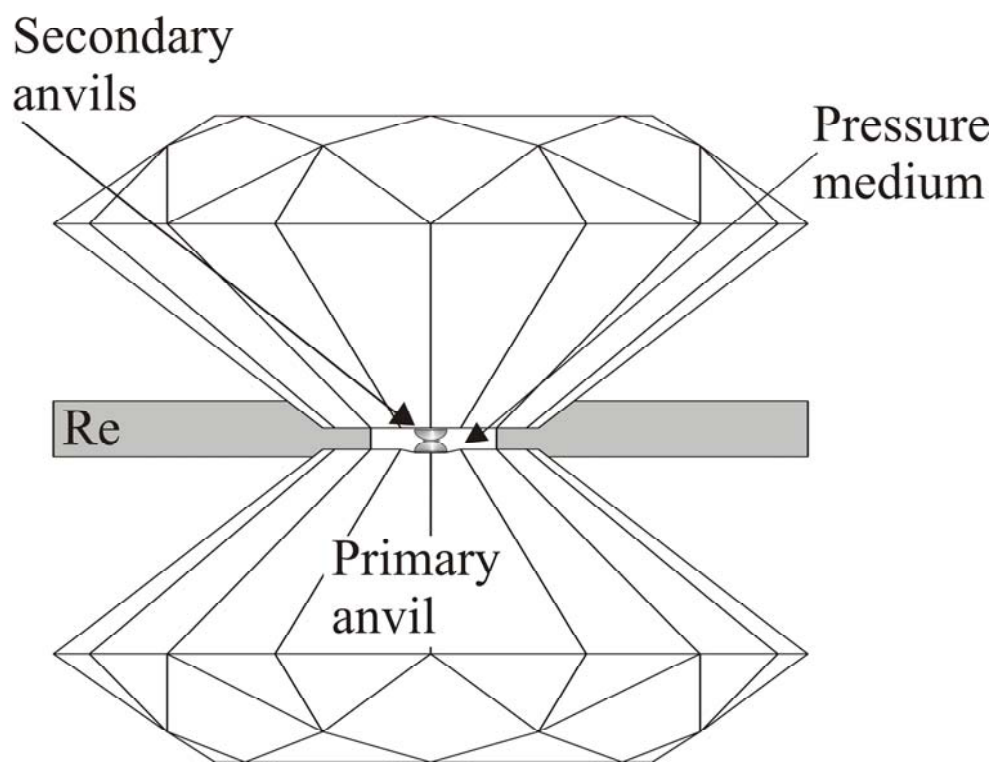

**Fig. S1. A schematic of the dsDAC.**

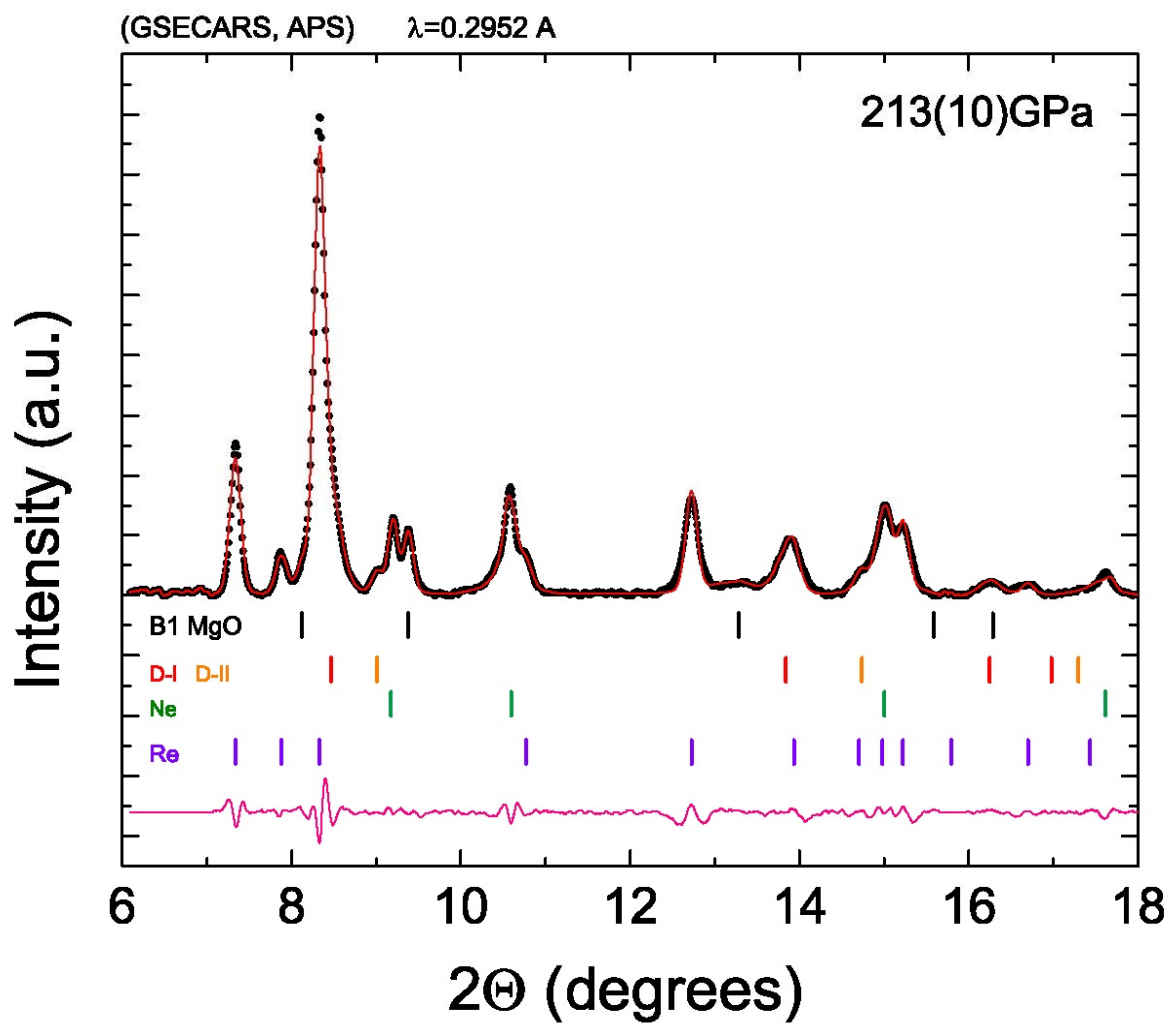

**Fig. S2. The diffraction pattern of MgO compressed in a gasketed dsDAC in a neon Ne pressure medium at the GSECARS beamline, APS.** Ticks correspond to: B1-MgO phase at 213(10) GPa (black ticks, noted "B1 MgO"); Re from the gasket at 50(3) GPa (purple ticks, noted "Re"); Ne pressure medium at 41(1) GPa (green ticks, noted "Ne"); outer part of the NCD anvil at 64(5) GPa (red ticks noted "D-I"), and inner part of NCD anvil at 204(10) GPa (orange ticks, noted "D-II"). Experimental data are shown by black dots; continuous red curve is the simulation using the full-profile (GSAS) software [15,16]; the difference of measured and calculated intensities is shown in magenta. The X-ray wavelength is 0.2952 Å.

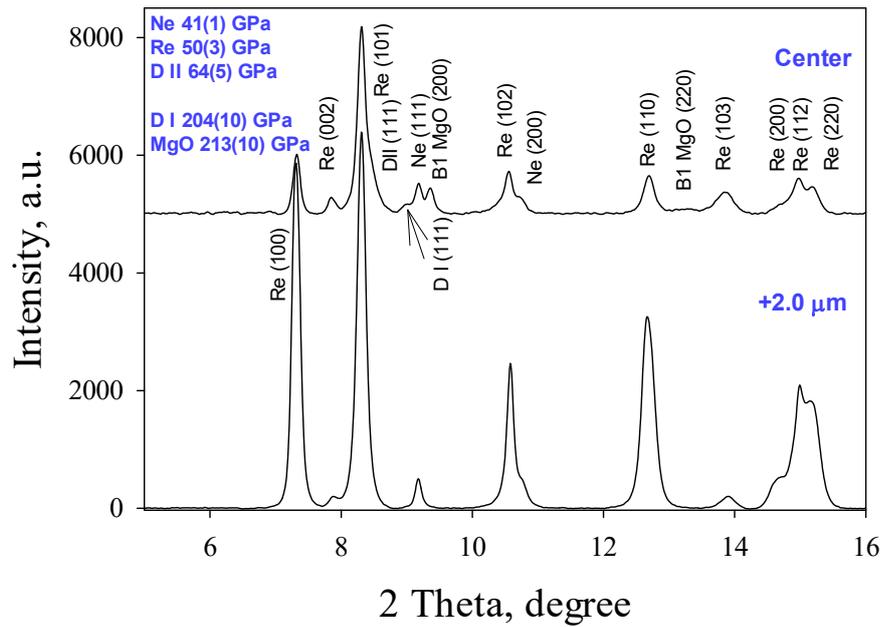

**Fig. S3. A comparison of the diffraction patterns obtained from the center of the sample in the gasketed ds-DAC and from a point which was 2 μm away.** The upper pattern (from the center) is the same as in Fig. S2. The lower pattern (2 μm away from the center) shows the diffraction only from Re gasket and Ne at about 45 GPa; MgO is not observed. The X-ray wavelength is 0.2952 Å (GSECARS, APS).

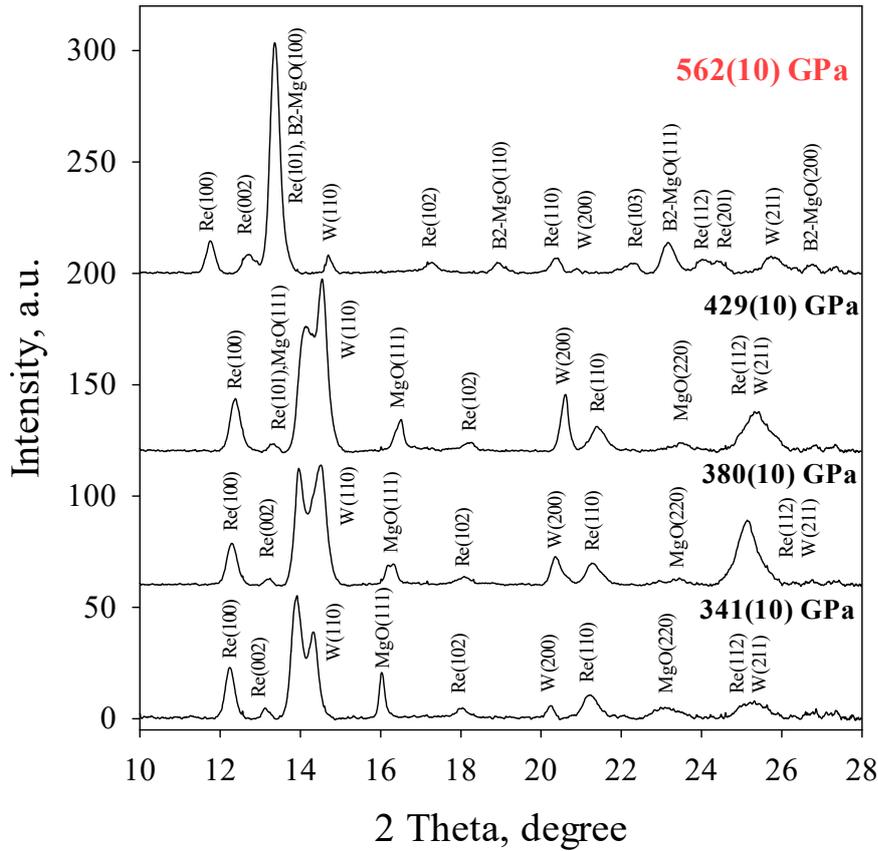

**Fig. S4. An example of the diffraction patterns of MgO compressed in gasketed dsDAC at ID16B beamline at the ESRF** (see Fig. 1). The X-ray wavelength is 0.4824 Å. The pressure indicated above each pattern corresponds to the pressure within the secondary pressure chamber, which was determined using the W EOS [12]. Upon compression, the increase of load on the primary bevelled anvils leads to the increase of pressure on the secondary Re gasket, which is located in the primary chamber of the dsDAC. For the three diffraction patterns, counting from the bottom of the figure, the pressure is equal to 79(3) GPa, 95(5), and 110(5) GPa, respectively, as determined from the EOS of Re [11]. This pressure is transmitted to a much smaller surface of the contact of the secondary anvils and is multiplied in the secondary chamber, which contains the MgO sample and W pressure calibrant, up to the values indicated in the figure. The pressure point 562(10) GPa became the last point in this experiment: after XRD measurements at the pressure point "429(10) GPa" had been made, upon further pressure increase, we observed formation of cracks on the bevel of one of the primary anvils, and the pressure on the Re gasket dropped to about 20 GPa, as determined from the Re EOS [11] (the uppermost diffraction pattern). However this start of failure of one of the primary anvils did not lead to immediate termination of the experiment, as the pressure around the secondary anvils increased to ~154 GPa (see Fig. S5), which led to the increase of pressure in the secondary chamber to 562(10) GPa.

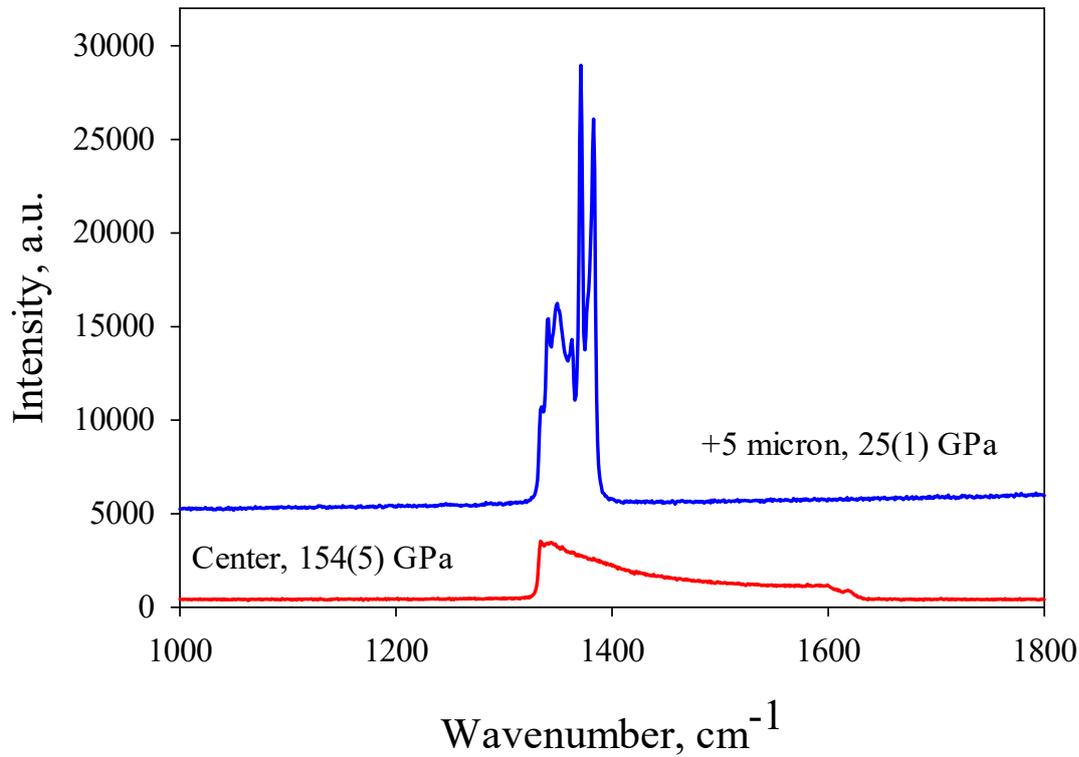

**Fig. S5. Raman spectra collected from the point close to the center of the 120-μm culet of the primary anvil (lower curve) and from the point of about 5 μm away from the center.** The spectra were taken during the experiment in the dsDAC at the ID16B at the ESRF after the pressurization of the dsDAC above 429(10) GPa (see Fig. S4). Using the position of the high-frequency edge of the diamond Raman band [22] we determined the pressure in the center of the primary anvil (lower spectrum) to be 154 GPa. This pressure was transmitted to the secondary anvils, although the pressure in the primary pressure chamber as a whole was preserved at the level of ~25 GPa (as determined by the diamond Raman shift method, upper spectrum).

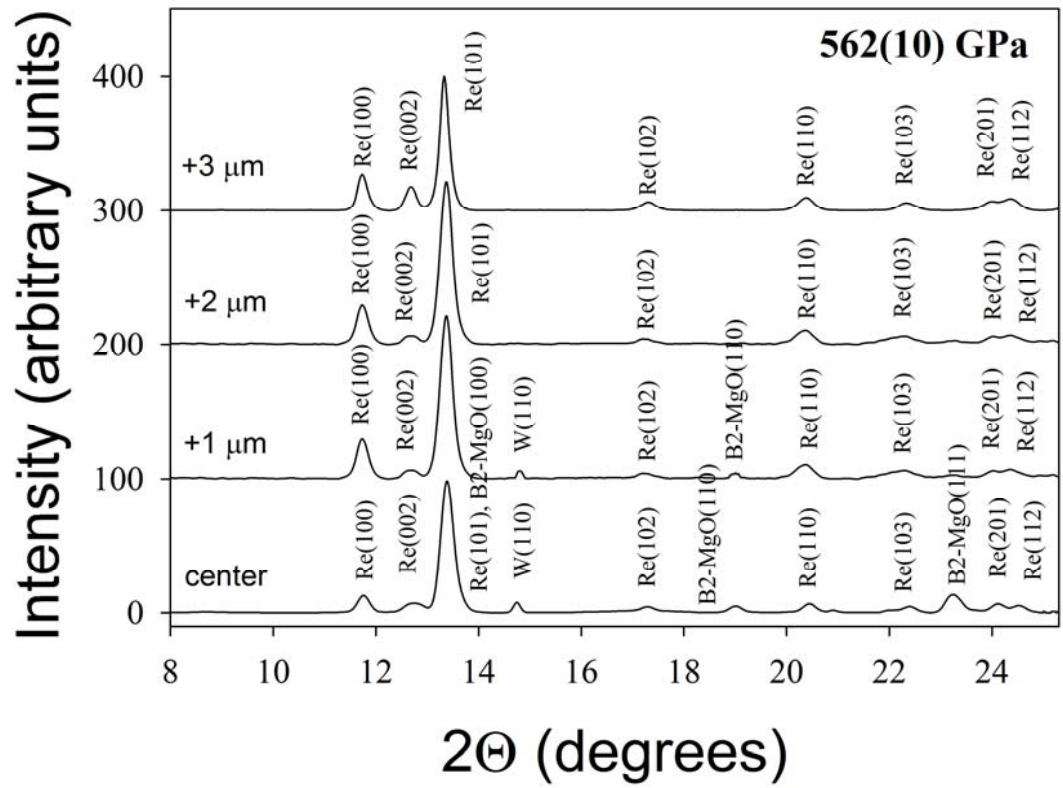

Fig. S6a

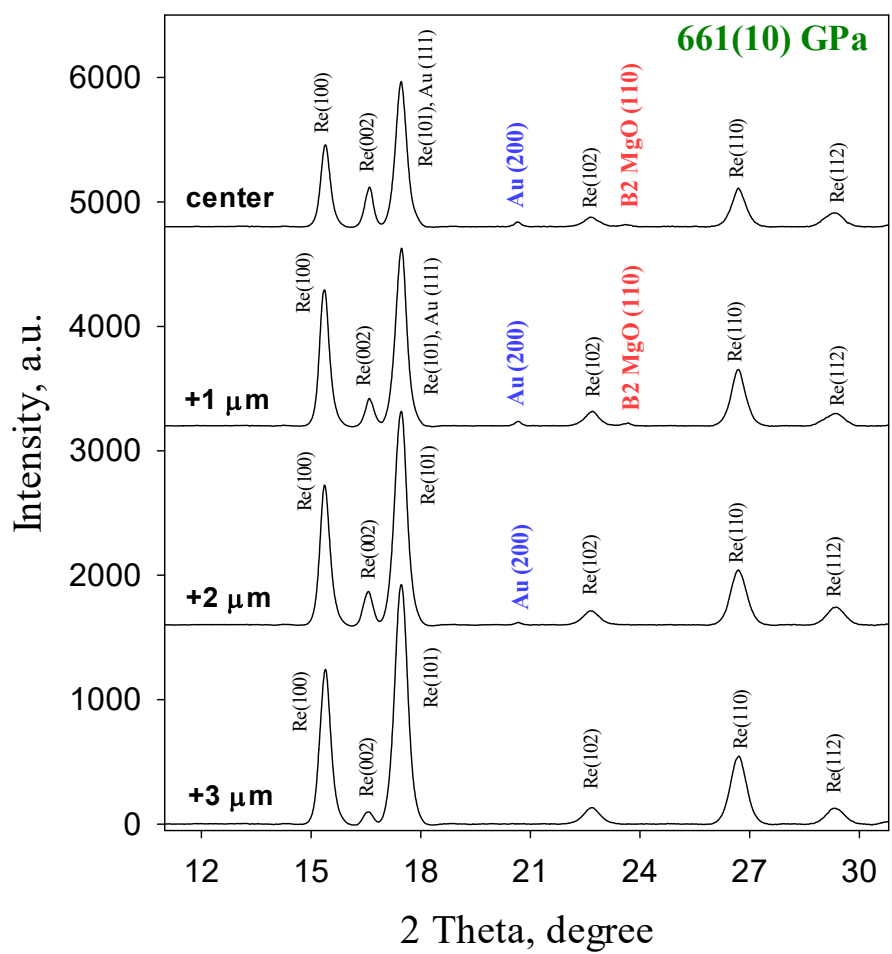

Fig. S6b

**Fig. S6. Examples of the diffraction patterns of the samples compressed in the gasketed dsDAC at (a) ID16 beamline (ESRF) and (b) P06 beamline (PETRA III) (see Figs. 1, 2).** One pattern was taken from the center of the sample; all other patterns were taken from the points away from the center with a step of 1 μm. X-ray wavelengths are (a) 0.4824 Å, (b) 0.5900 Å. No pressure gradients are observed within the samples, but most probably this can be explained by the large tails of the X-ray beam at each of the beamlines.

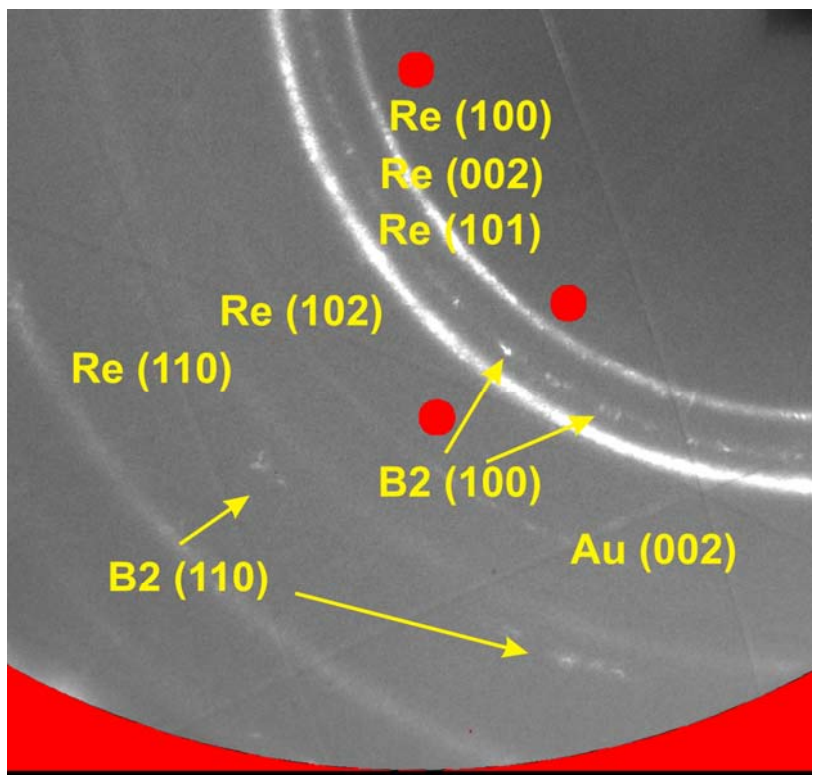

Fig. S7a

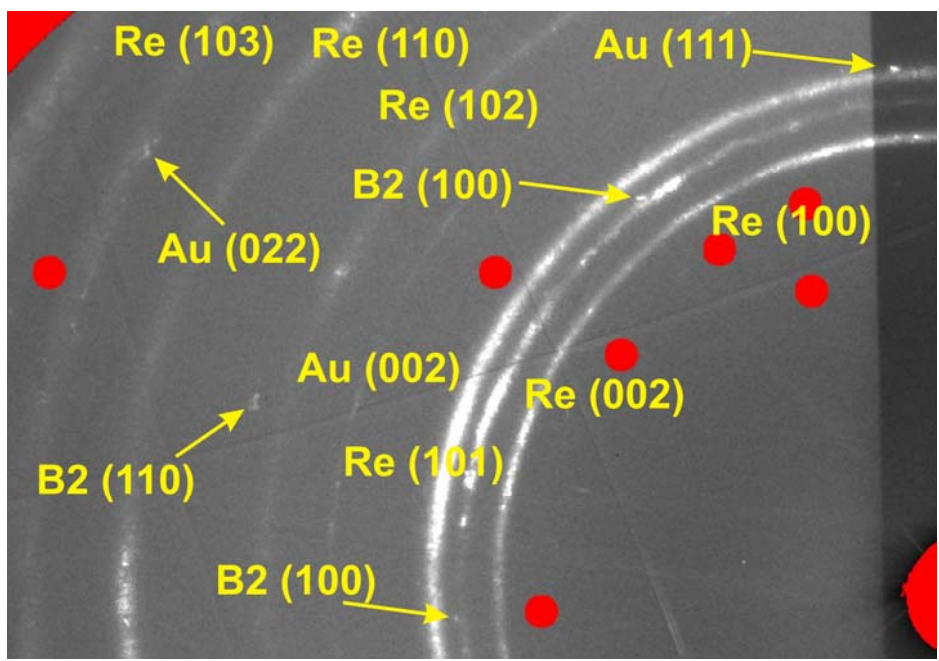

Fig. S7b

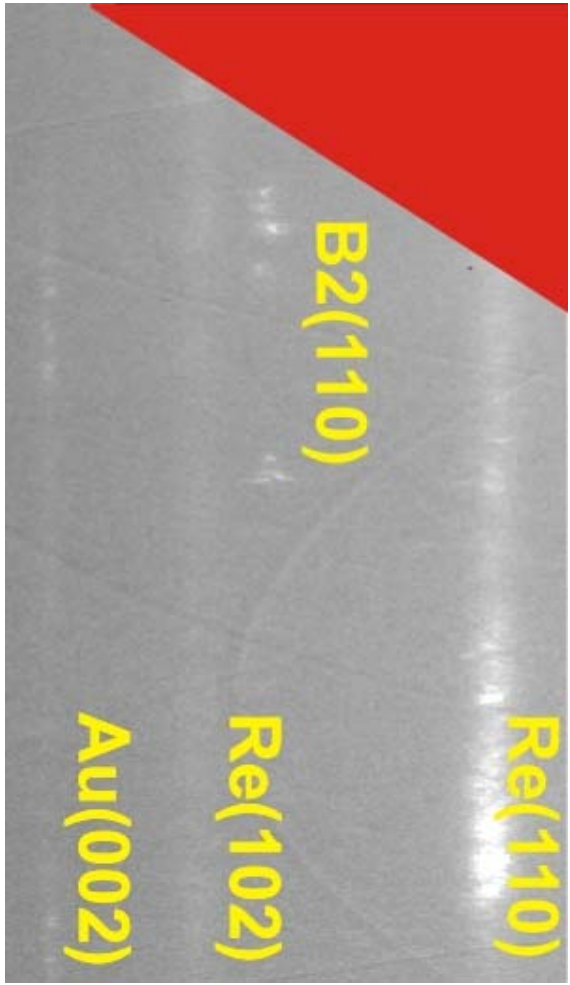

Fig. S7c

**Fig. S7. Parts of the 2D diffraction images: (a) and (b) as collected and (c) "caked") for a sample compressed in the gasketed dsDAC at P06 beamline, PETRA III.** The diffraction was collected at 661(10) GPa (see Fig. S6). Red areas mark the parts of the detector, which were blocked by the body of the DAC; red dots mask the artefacts related to the detector.